\documentclass[preprints,review,accept,moreauthors]{Definitions/mdpi} 

\firstpage{1} 
\pubvolume{xx}
\issuenum{1}
\articlenumber{5}
\pubyear{2019}
\copyrightyear{2019}
\history{Received: 12 July 2019 ; Accepted: 16 August 2019}

\Title{The Wolf-Rayet Content of the Galaxies of the Local Group and Beyond}

\Author{Kathryn Neugent $^{1,2}$\orcidA{} and Phil Massey $^{2,3}$\orcidB{}}

\AuthorNames{Kathryn Neugent and Phil Massey}

\address{%
$^{1}$ \quad University of Washington; kneugent@uw.edu\\
$^{2}$ \quad Lowell Observatory; massey@lowell.edu\\
$^{3}$ \quad Northern Arizona University}

\abstract{Wolf-Rayet stars (WRs) represent the end of a massive star's life as it is about to turn into a supernova. Obtaining complete samples of such stars across a large range of metallicities poses observational challenges, but presents us with an exacting way to test current stellar evolutionary theories. A technique we have developed and refined involves interference filter imaging combined with image subtraction and crowded-field photometry.  This helps us address one of the most controversial topics in current massive star research: the relative importance of binarity in the evolution of massive stars and formation of WRs.  Here we discuss the current state of the field, including how the observed WR populations match with the predictions of both single and binary star evolutionary models. We end with what we believe are the most important next steps in WR research.}


\begin{document}

\section{Wolf-Rayet Star Primer}
Wolf-Rayet (WR) stars are hot, luminous stars whose spectra are dominated by strong emission lines, either of helium and nitrogen (WN-type) or helium, carbon, and oxygen (WC and WO type).  It is generally accepted that these are the He-burning bare stellar cores of evolved massive stars \cite{Meynet05}. Mass loss (whether from binary interactions or stellar winds) first strips away the outer layers of a massive star to reveal the products of CNO hydrogen-burning, nitrogen and helium,  creating a nitrogen-rich WR (WN-type). If enough subsequent mass loss occurs, these layers are then stripped away, revealing the triple-$\alpha$ helium-burning products, carbon and oxygen, creating a WC star. Further evolution and mass loss may result in a rare-type oxygen-rich WR (WO-type). 

The mass loss that shapes the evolution of these stars can occur through two main channels: binary and single-star evolution. The relative importance of each method is still one of the most important questions facing massive star evolution today. In a binary system, the more massive star will expand first and be stripped by the companion star, revealing the bare stellar core of a WR. In single star evolution, the star will follow the Conti scenario \cite{Conti75,Maeder94}. In the Conti Scenario, stars with initial masses greater than $\sim 30M_{\odot}$ will form on the main-sequence as massive O-type stars. As they evolve, the stellar winds will continue to strip more and more material from their surfaces until they first turn into WNs, and then (depending on the strength of the stellar winds), WCs and possibly WOs. Stars with initial masses greater than $85M_{\odot}$ will also briefly pass through the turbulent Luminous Blue Variable (LBV) phase, shedding material that way.

Single-star evolution is highly dependent on the strength of the stellar-wind mass-loss rates, which are in turn dependent on the metallicity of the birth environment. Since this mass-loss is driven by radiation pressure on highly ionized metal atoms, a massive star born in a higher metallicity environment will have a higher mass-loss rate, and thus the mass limit for becoming a WR would be lower in a higher metallicity environment. If stellar winds dominate the mass-loss mechanism (as opposed to binary evolution), it follows that WC stars will be more common relative to WN stars in high metallicity galaxies while low metallicity galaxies will have few or even no WCs. It also follows that, assuming only single-star evolution, WOs will be rare in all except the highest-metallicity galaxies. Thus the presence of WOs in a low-metallicity environments (as we discuss later) suggests that binary-evolution plays an important role in the creation and evolution of WRs in at least some cases \cite{Moffat1990,BPASS2.1}.  Or, as  J. J. Eldridge and collaborators have put it \cite{BPASS2.1}, "Single-star stellar winds are not strong enough to create every WR star we see in the sky."  

Determining the relative number of WC-type and WN-type WRs (the WC to WN ratio) allows us to test stellar evolutionary models by comparing what we see observationally to what the models predict as they scale with the metallicity of the environment. Reliable evolutionary tracks affect not only the studies of massive stars, but the usefulness of population synthesis codes such as STARBURST99 \cite{Leitherer1999}, used to interpret the spectra of distant galaxies. For example, the inferred properties of the host galaxies of gamma-ray bursts depend upon exactly which set of stellar evolutionary models are included \cite{Levesque2007}. It is also important for improving our knowledge of the impact of massive stars on nucleosynthesis and hence the chemical enrichment of galaxies \cite{Meynet2008}. Thus, determining an accurate ratio of WC to WN stars in a galaxy turns out to have its uses far beyond the massive star community \cite{Maeder1980}.  Additional diagnostics include the relative number of red supergiants (RSGs) to WRs, and the relative number of O-type stars to WRs.

The galaxies of the Local Group provide an excellent test-bed for such comparisons between the observations and models because they allow us to determine a {\it complete} population of different types of stars. In all except the most crowded of regions (such as 30 Doradus in the Large Magellanic Cloud), stars can be individually resolved by ground-based telescopes and instruments. Such photometric studies have been done previously (such as the Local Group Galaxy Survey [LGGS]\cite{Massey2002}), but photometry alone can't be used to detect Wolf-Rayet stars. Thus, as we will discuss in this article, other methods such as interference filter imaging and image subtraction must be employed. The WR-containing galaxies of the Local Group span a range in metallicity from 0.25$\times$ solar in the Small Magellanic Cloud (SMC) \cite{Russell1990} to 1.7$\times$ solar in M31 \cite{NelsonM31}. This allows us to compare the observations against the model predictions across a large range of metallicities, which is important given the strong dependence on stellar evolution to mass-loss rate. Thus here we focus our discussions on WRs in the galaxies of the Local Group.

In this review paper we will first discuss how WRs were found in the past as well as current methods. We'll review the current WR content of the Local Group Galaxies and Beyond while discussing a few important and surprising findings made along the way. Next we'll discuss the important issue of binarity and how it influences the evolution of WRs. Finally, we'll describe how to obtain the physical parameters of such stars using spectral modeling programs before ending with a discussion of how the evolutionary models compare to our observed number of WRs.

\section{Surveys for Wolf-Rayet Stars}
\subsection{The Milky Way}
The first survey for Wolf-Rayet stars (inadvertently) began in 1867 when Charles Wolf and Georges Rayet were examining spectra of stars in Cygnus using a visual spectrometer on the 40-cm Foucault telescope at the Paris Observatory.  They came across three very unusual stars.  While the spectra of most stars are dominated by absorption lines, these stars had mysterious strong, broad emission lines.  (These stars were later designated and classified as HD 191765, WN5; HD 192103, WC8;  and HD 192641, WC7.) 

\begin{figure}[H]
\centering      
\includegraphics[width=0.4\textwidth]{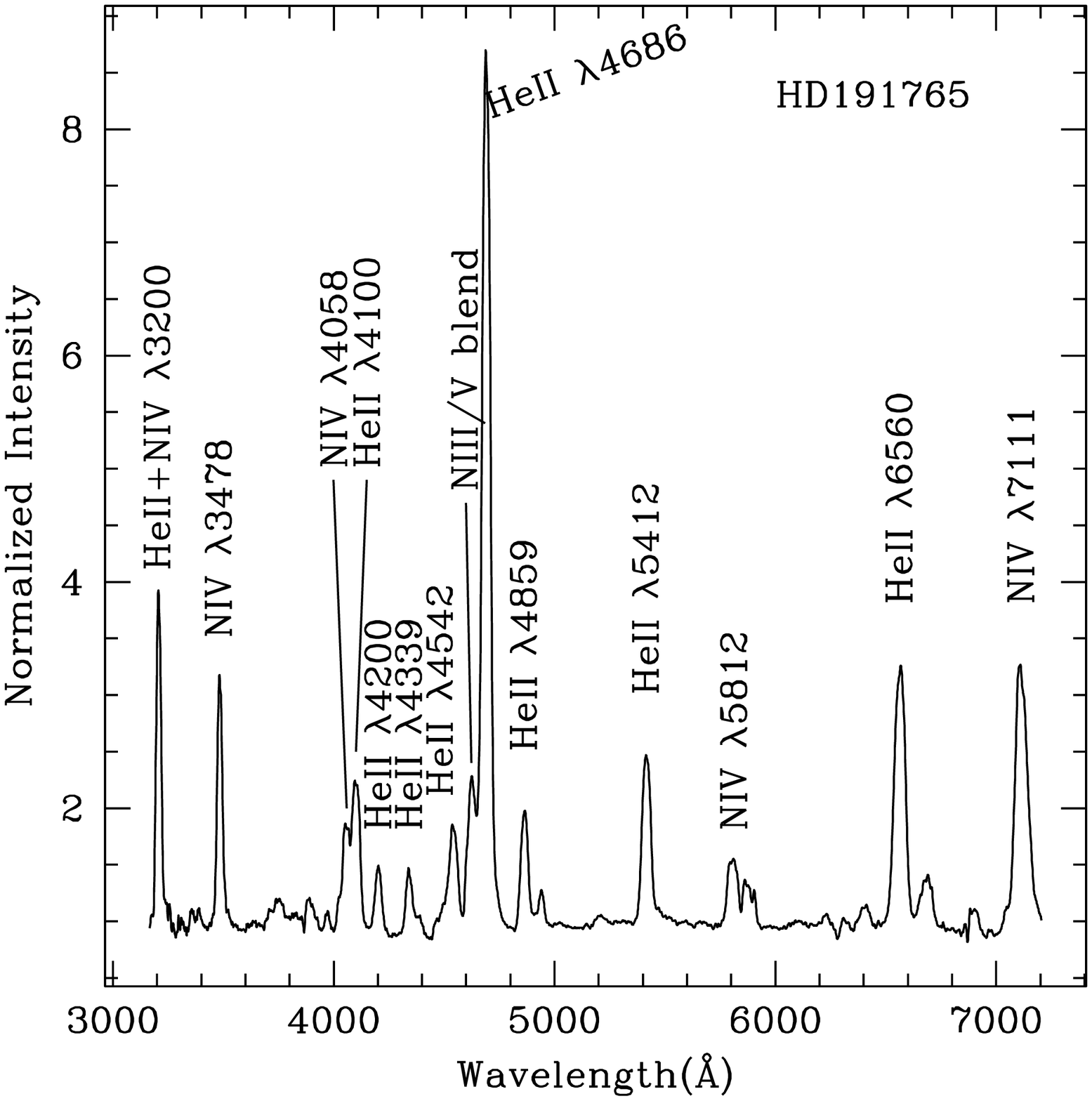}
\includegraphics[width=0.4\textwidth]{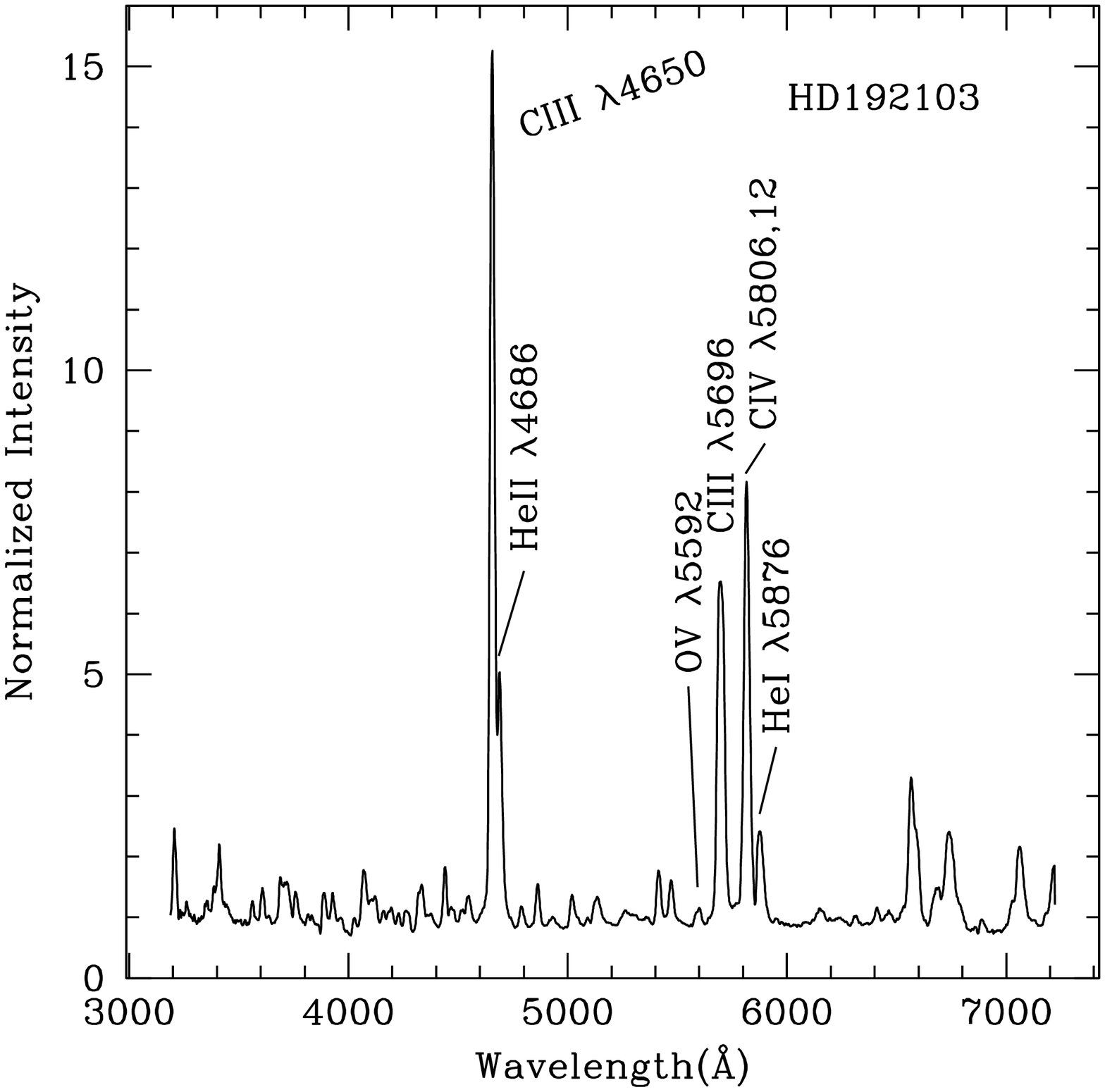}
\caption{\label{fig:WRspec} The spectra of two of the first discovered WR stars. Left: HD 191765 is a WN star, with unusually broad and strong lines. Its classification is a "WN5" subtype.  Right: HD 192103 is a WC star, with a "WC8" subtype.}
\end{figure}

The correct identification of the spectral features was lacking for nearly 60 years after their discovery: it was Carlyle Beals, a Canadian astronomer, who correctly identified the lines as due to ionized helium, nitrogen, and carbon \cite{Beals30}.
The width of these lines were understood as being due to Doppler broadening of thousands of km s$^{-1}$, a result of the outflow rates of the strong stellar winds in the formation region of these lines \cite{Beals29,Cass75,Conti79}. Example spectra are shown in Figure~\ref{fig:WRspec}.

WN-type Wolf-Rayet stars are further classified primarily by the relative strengths of  N\,{\sc iii} $\lambda$4634,42, N\,{\sc iv} $\lambda$4058, and N\,{\sc v} $\lambda$4603,19, while the classification of WC-type WRs is based upon the relative strengths of O\,{\sc v} $\lambda$5592, C\,{\sc iii} $\lambda$5696, and C\,{\sc iv} $\lambda$5806,12.  The system was first proposed by Lindsey Smith \cite{Smith68}, although some extension to earlier and later types of WNs have been made by others \cite{vanderHuchtVI,vanderHuchtVII}; a classification scheme for WO stars was developed by Paul Crowther and collaborators \cite{WOs}.  As with normal stars, a lower number is indicative of higher excitation, i.e., WN2 (hotter) vs.\ WN9 (cooler), WC4 (hotter) vs.\ WC9 (cooler), WO1 (hotter) vs.\ WO4 (cooler).  

The late-type WNs are morphologically similar to O-type supergiants, known as "Of-type" type stars, in that the latter show N\,{\sc iii} $\lambda$4634,42 and He\,{\sc ii} $\lambda$4686 emission, also the result of stellar winds.  The late-type WNs are more extreme, however, with stronger lines.  In general, WNs (and WRs in general) do not show absorption lines; rather, all of the lines are formed in the stellar winds.  There are, however, exceptions, such as HD 92740, a singled-lined WR binary in which the emission and absorption move together in phase \cite{ContiNiemela}.   It was the similarity between Of-type and WNs that led in part to the Conti scenario \cite{Conti75}. 

As summarized in \cite{MasseyThesis}, a total of 52 similar stars were discovered by Copeland, Fleming, Pickering, and Respighi in the 25 years that followed Wolf and Rayet's discovery.  These findings, and early visual work by Vogel in 1885, and photographic studies of their spectra by Pickering in 1890, are discussed in the contemporary review by Julius Scheiner and Edwin Frost in their 1894 publication  {\em A Treatise on Astronomical Spectroscopy} \cite{ScheinerFrost}.
William Campbell (who served as director of Lick Observatory 1901-1930) published the first catalog of these 55 Galactic WRs in 1894 \cite{Campbell}.  Additional WRs were discovered as by Williamina Fleming, Annie J. Cannon and coworkers as part of the Henry Draper catalog project, and accidental discoveries continued to be made over the years.
The first modern catalog of Galactic Wolf-Rayet stars compiled by Karel van der Hucht and collaborators in 1981 \cite{vanderHuchtVI}.  Titled "The VIth Catalog" (Campbell's was considered the first), the work included extensive bibliographies and references to earlier studies.  This catalog contained 168 WRs. The next edition, in 2001 \cite{vanderHuchtVII}, listed 227 spectroscopically confirmed Galactic WRs,  with an "annex" in 2006 \cite{vanderHucht06} bringing the number known to 298.  The most-up-to-date catalog of Milky Way WRs is maintained on-line by Paul Crowther (http://www.pacrowther.staff.shef.ac.uk/WRcat/), which contained 661 entries as of of this writing, June 2019.

Systematic searches for WRs in the Milky Way are stymied by the vast angular extent that needs to be examined (the entire sky!), and variable and sometimes high reddening.  The Henry Draper catalog is probably complete down to an apparent magnitude of 10th or 11th, except in regions of crowding.  Spectroscopic surveys of young clusters or OB associations reveal additional WR finds yearly; possibly the most extreme example is that of Westerlund 1 and various open clusters near the Galactic Center; see \cite{vanderHucht06} and references therein. However, the large increase in the number of WR stars known in the Galaxy in the past 15 years has has come about primarily from the use near- and mid-IR colors to identify WR candidates \cite{2007MNRAS.376..248H, 2011AJ....142...40M, 2009PASP..121..591M, 2011AJ....142...40M, 2014AJ....147..115F, 2018MNRAS.473.2853R}, a method first pioneered by Schuyler van Dyk and Pat Morris, plus the use of narrow-band IR imaging in the K-band \cite{2009AJ....138..402S, 2012AJ....143..149S}, pioneered by Mike Shara.  Optical or near-IR spectroscopy is then used to confirm the color-selected candidates.

With the advent of {\it Gaia}, it is now possible for the first time to actually derive distances to many of these Wolf-Rayet stars. However, difficulties of constructing meaningful volume-limited samples remain for Galactic studies.  As discussed later, WN-type WRs are harder to find than WC-type due to their weaker lines; at the same time, WC stars may be dustier (and thus fainter) than WN stars in the same location.  They also cover a limited range in metallicity compared to what can be achieved by using the non-MW members of the Local Group. Finally, observations of Galactic WRs may be more difficult due to reddening than those in much further, but less reddened, regions.  Thus, Galactic studies still have limited value for testing models of stellar evolution theory.   Thus for the rest of this review, we will focus on the WR content of galaxies outside our own.

\subsection{Early Searches for Extra-Galactic WRs}

\subsubsection{Large Magellanic Cloud} 
As part of the Harvard spectral surveys, Anne J.\ Cannon and Cecilia Payne (later Payne-Gaposchkin) identified 50 Wolf-Rayet stars in the Large Magellanic Cloud (LMC) according to Bengt Westerlund \& Alexander Rodgers (1959) \cite{West59} quoting an early review article on the stellar content of the LMC by Gerard de Vaucouleurs and collaborators \cite{deVacLMC}.  Westerlund \& Rodgers carried out their own search of the LMC,  the first systematic search for WR stars in another galaxy, using slitless (objective prism) spectroscopy to identify 50 WRs, 36 of which were in common with the Harvard studies \cite{West59}. They note that nine Harvard O-type stars in the 30 Doradus region had been recently reclassified as WN by Michael Feast and coworkers \cite{Feast} in the previous year. Two decades later, Marc Azzopardi \& Jacques Breysacher (1979) completed an even more powerful objective prism survey using an interference filter to further reduce the effects of crowding \cite{AzzoBrey79LMC}. This increased the number of known WRs in the the LMC to 100. Accurate spectral types of these 100 LMC WRs were subsequently published by Breysacher in 1981 \cite{Brey81}. In that paper, Breysacher estimated that the LMC likely contained a total of 144 $\pm$ 20 LMC WRs, with 44 left to be discovered. He further speculated that the majority of these undiscovered WRs would be found deep within the cores of dense H II regions where slitless spectroscopy often fails. (Indeed, the "final census" catalogue of LMC WRs, discussed below, lists 154 separate WRs \cite{Neugent18}, well within Breysacher's estimate of 144 $\pm$ 20.)  These early studies culminated in Breysacher's et al.'s "Fourth Catalog" of LMC WRs \cite{BAT99} (hereafter BAT99), which listed 134 LMC WRs.

The R136 cluster merits separate attention, as investigations of its stellar content led to the recognition that not all luminous stars with WR-like spectra are evolved objects.  R136 is of course the central object at the heart of the 30 Doradus nebula in the LMC.  Once thought to house a supermassive star, early {\it Hubble Space Telescope (HST)} images showed it was even more interesting, the core of a super star cluster, with over 3500 stars (120 of which are blue and more luminous than $M_V\sim -4$) most of which lie within 8" (2pc) of the semistellar R136 cluster \cite{Hunter95}.  Using ground-based spectroscopy in 1985, Jorge Melnick had identified 12 WR stars in or near the central cluster \cite{Melnick85}.  When Deidre Hunter and collaborators analyzed the first {\it HST} images of the cluster in 1995, this created a conundrum: the isochrones indicated that the lower mass stars had ages of only 1-2~Myr, but the presence of WR stars implied ages for the massive stars of 3-4~Myr \cite{Hunter95}. Why had the formation of high mass stars, with their strong stellar winds, not stopped star formation in the cluster?  Melnick had also found early-type O stars in the cluster, possibly as early as O3, although the presence of strong nebulosity made this classification uncertain, and this also seemed to conflict with the ages of the WR stars, as the O3 phase lasts for only a million years. Massey and Hunter obtained {\it HST} spectroscopy of 65 of the hottest, bluest stars in the cluster, and discovered two amazing facts: (1) the vast majority of these stars were of O3, and that (2) the WR stars were not common, garden-variety WNs \cite{MasseyHunter}.  Rather, they were 10$\times$ more luminous in the V-band than normal WRs,  and their spectra were still rich in hydrogen.  Massey and Hunter argued that a similar situation existed in the Galactic giant H\,{\sc ii} region NGC 3603, where both O3 stars and WRs were known \cite{Drissen95}; they examined the archival spectra and concluded that those WR stars were like the H-rich super-bright WR in R136.  The obvious conclusion was that these were young (1-2~Myr) objects still burning hydrogen whose high luminosities simply resulted in WR-like emission features, in essence, Of-type stars on steroids \cite{MasseyHunter}. This interpretation built on the important result the previous year by Alex de Koter and collaborators who found that one of the over-luminous, hydrogen-rich WR stars in the core of the R136 cluster had a normal hydrogen abundance, and who had originally suggested that this and similar were still in the hydrogen-burning phase \cite{deKoter}.

\subsubsection{The Small Magellanic Cloud}
The identification of WRs in the SMC followed a similar pattern, but thanks to its smaller angular size compared to the LMC, a complete census became possible earlier than for the galaxies discussed above.  As summarized in an earlier review \cite{MasseyRev13},  four WRs had been found by general spectroscopic studies \cite{Brey78} when Azzopardi \& Breysacher used the same technique of objective prism and interference filter photography to find four additional WRs, bringing the total up to eight \cite{AzzoBrey79SMC}.  A ninth WR was found by spectroscopy from objective prism photography \cite{Morgan}.  In 2001, Massey and summer student Alaine Duffy carried out the first CCD survey for WRs in the  SMC \cite{Massey01}. They used on-band, off-band interference filter imaging campaign with the wide-field CCD camera on the CTIO Curtis Schmidt to cover most of the SMC. Photometry of 1.6 million stellar images helped identify a number of candidates, including all of the known SMC WRs, at high significance levels.  Two new WNs were then confirmed by follow-up spectroscopy, bringing the total to 11. The survey also found a number of Of-type stars, demonstrating that the survey was sensitive to even the weakest-lined WNs.  However, shortly after this a 12th WR star was discovered in the SMC \cite{Massey12th}. This star had been too crowded to have been found in the Massey \& Duffy survey. Of these 12 WRs, 11 are of WN-type and only 1 is of WC-type. (Actually the strength of O~VI lines qualifies this as a WO-type star \cite{WOs}.) This low WC/WN ratio is consistent with our expectations based upon the SMC's low metallicity.

Quantitative studies of the strength of He\,{\sc ii} $\lambda$4686 emission in SMC WN stars by Peter Conti and collaborators \cite{Conti89} showed that the line was weaker than in WNs of similar types in the Milky Way or LMC, also consistent with the expectation that stellar winds would be weaker in lower-metallicity environments.

\subsubsection{Beyond the Magellanic Clouds}

The first WR stars to be discovered beyond the Magellanic Clouds were in the nearby spiral galaxy M33.
In 1972 James Wray and George Corso pioneered the interference-filter method of searching for WRs by comparing images of M33 taken through an interference filter centered on the C\,{\sc iii} $\lambda$4650 and He\,{\sc ii} $\lambda$4686 emission complex with that of a continuum image \cite{WrayCorso}.  WR candidates would stand out by being brighter in the on-band compared to non-WR stars in the field. Their paper contained spectroscopic confirmation of two of their 25 candidates (thanks to Roger Lynds); both stars were of WC-type, although Lindsey Smith is quoted as saying that the spectra were "not quite like any I have seen from either the Galaxy or the Magellanic Clouds." (This was probably more due to the poor quality of these early spectroscopic efforts on these faint objects, which pushed the limits of photographic spectroscopy at that time.)  Spectroscopy of three other candidates followed five years later by Alex Boksenberg, Allan Willis, and Leonard Searle using one of the first digital photon-counting systems \cite{Boksenberg}.  A search using photographic "grism" imaging on the Kitt Peak 4-meter (a technique similar to objective prism survey but using a grating prism and a much larger telescope) carried out by Bruce Bohannan, Conti, and Massey revealed a host of H\,{\sc ii} regions in M33, but only five more WRs \cite{Bohannan}.  Spectroscopy of the stars in M33's H\,{\sc ii} regions by Conti and Massey in 1981 was more effective, identifying 14 more WRs \cite{Casual}; some were in common with the nearly contemporaneous study of the stellar content of NGC 604, the largest H\,{\sc ii} region in M33, by Mike Rosa and Sandro D'Odorico \cite{Rosa,Dod300}.   The properties of some of these stars were highly unusual, with higher luminosities and more hydrogen than normal WR stars, similar to what would be eventually noted in the R136 cluster as mentioned above.  A photographic search with the 3.6-meter Canada-France-Hawaii telescope with followup spectroscopy on the Kitt Peak 4-meter provided the first galaxy-wide survey, including 41 newly found WRs \cite{MasseyConti83}.
This 1983 Massey \& Conti catalog included all previous known WRs, for a total of 79 WRs, and revealed a trend in the relative number of WCs to WNs as a function of galactocentric distance within M33.  Quantitative analysis of the lines (measurements of line strengths and widths) and absolute magnitudes showed no gross differences between the M33 WRs and those of the Milky Way or Magellanic Clouds \cite{MasseyConti83,MCA}, refuting the Smith's first impression from the Lynds' earlier spectroscopy.

The first use of CCDs to survey for WRs was carried out by Taft Armandroff and Massey in 1985 using the newly implemented prime-focus CCD camera on the Cerro Tololo Blanco 4-meter telescope \cite{AM85}.  They had refined the interference-filter method to include a three-filter system, with one centered on C\,{\sc iii} $\lambda$4650, another on He\,{\sc ii} $\lambda$4686, and a third on neighboring continuum, and used these with a CCD to search for WRs in the dwarf galaxies IC 1613 and NGC 6822, as well as two M33 test fields.  One WR star had been previously identified in IC 1613, a WC star (now considered a WO) discovered in an H\,{\sc ii} region by D'Odorico and Rosa in 1982 \cite{DOd}, and subsequently studied by Kris Davidson and Tom Kinman \cite{DavidsonKinman}.  Similarly a WN-type WR had previously been found in NGC~6822 by Westerlund and coworkers using an objective prism \cite{WesterlundN6822}.  These early CCDs were incredibly tiny compared to what are in use today, and multiple fields were needed to cover even these relatively small galaxies.  These CCDs were also incredibly noisy (with read-noise of 100 e- compared to typically 3 e- today).  Armandroff and Massey found 12 "statistically significant" WR candidates in NGC~6822 and 8 in IC~1613. However, only 4 of the NGC~6822 WR candidates proved to be real (including the one that was previously known), and the only IC~1613 WR candidate that checked out was the one already known \cite{AM91}.

A search for WR stars in the dwarf galaxy IC~10 proved the most surprising of any of these early studies.
Despite its small size, 16 WR candidates were initially found by Massey, Armandroff, and Conti \cite{MAC92}, 15 of which were quickly confirmed \cite{AM95}, causing the authors to recognize this as the nearest starburst galaxy. Despite the galaxy's low metallicity, the relative proportion of WC stars was very large.  Was this suggestive of a top-heavy initial mass function as has been historically suggested for other starbursts \cite{Rieke}, or is indicative that an even larger number of WRs (predominantly WN) remained to be discovered, as suggested by \cite{MasseyHolmes}?  This issue is still not settled.  The current count is 29 spectroscopically confirmed WRs \cite{Tehrani}, with additional candidates still under investigation.

The situation for M31 was probably the worst.  Interference photography by Tony Moffat and Mike Shara identified a few of the strongest-lined WRs \cite{MS83,MS87}; CCD imaging through interference filters 
by Massey and collaborators went much deeper but covered only a small portion of the galaxy \cite{MAC86,AM91}. 

These early studies culminated in the 1998 paper by Massey and Olivia Johnson \cite{MJ98}, who identified additional M33 WR stars found using a larger- format (and less noisy) CCD, and provided a catalog of all of the known extragalactic WR stars beyond the Magellanic Clouds.  For the purposes of this review, we will consider that the end of the "early era" of WR searches.   Although completeness indeed would prove to be a problem, the following facts had emerged:
\begin{itemize}
\item The WC/WN ratio appeared to be strongly correlated with metallicity, with the exception of the starburst galaxy IC~10.
\item Late-type WC stars (WC7-9) were found only in regions of high metallicity, while WCs in low-metallicity regions were invariably of early type (WC4s).
\item The spectral properties of a given WR type were generally similar regardless of the environment, although weaker emission is found in the WNs of lower metallicity, indicative of smaller mass-loss rates.
\item Giant H\,{\sc ii} regions (NGC 604, 30 Dor, NGC 3603) contained very luminous stars whose spectra showed WR-like features, but which were hydrogen-rich.  These stars were basically "super Of-type stars," stars that are so massive and luminous that their atmospheres are extended creating WR-like features but which are likely still hydrogen-burning objects.
\end{itemize}

\subsection{Motivation for New Studies}

\begin{figure}[H]
\centering      
\includegraphics[width=0.5\textwidth]{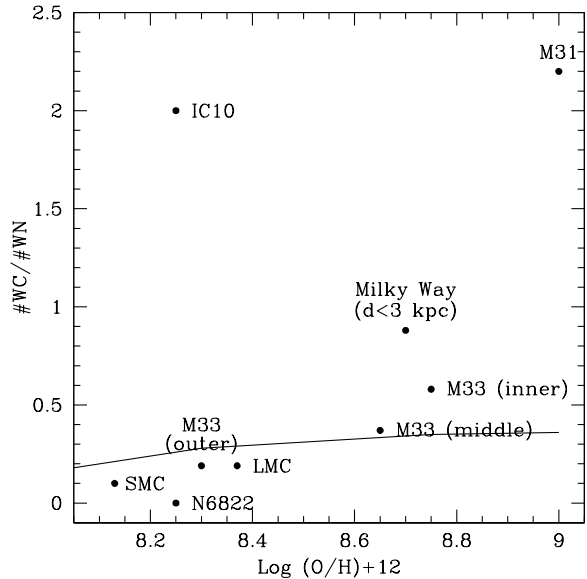}
\caption{\label{fig:Geneva2005} The state of our knowledge of the WC/WN ratio vs metallicity in the mid 2000s. The points are from the 1998 Massey \& Johnson summary \cite{MJ98}. The solid curve shows the predictions based upon the 2005 Geneva evolutionary models that included rotation for the first time \cite{MM05}.   Note that while both show an increase in the WC/WN ratio with metallicity, there is a large discrepancy between the observed results and model predictions at higher metallicity values. Recall that NGC~6822 contains only 4 WRs (all of WN-type) and the SMC only 12 WRs (one of which is a WC/WO); thus deviations from the models for these two galaxies is not significant.
}
\end{figure}

As of the early 2000s, our knowledge of the LMC's WR population was thought to be relatively complete thanks to the work of Breysacher's BAT99 catalog \cite{BAT99}. However, other galaxies of the Local Group, namely M31 and M33, still lacked galaxy-wide surveys. Figure~\ref{fig:Geneva2005} shows the observed WC/WN ratio compared to the 2005 Geneva Evolutionary Group's model predictions \cite{MM05}. (These were the first complete set of models at different metallicities which included the important effect of rotation.)  Notice first that the observed relative number of WCs to WNs increases with metallicity. This is exactly what we would expect given single-star evolution because higher metallicity environments will allow more WCs to form. This increase in ratio vs.\ metallicity is additionally what the models predict. However, a comparison between the models and the observations show that the relative number of predicted WRs is not consistent between the two. Additionally, the models do a particularly poor job of predicting the WC to WN ratio at higher metallicities, such as in M31 and M33.  

Clearly a problem existed -- but was it a failing of the models or observations (or, both)? Given the complexities of modeling the physics at the end of a massive star's life, it made sense that there could be some deficiencies in the models. However, there were a few reasons that suggested that the observations were actually at fault. For one, as discussed above, there was still no galaxy-wide targeted survey of WRs in the LMC, M31 or M33; only the SMC had been well covered by the Massey \& Duffy survey.  The vast majority of WRs that had been discovered within those galaxies had been discovered either by accident or as part of a survey of a limited portion of the galaxy.  Additionally, crowding of tight OB associations (where we expect to find the vast majority of WRs) makes finding even bright, strong-lined WRs difficult. Thus, telescopes with more resolving power could help disentangle the tightly-packed regions. Finally, and perhaps more importantly, there is a strong observational bias towards detecting WC-type stars over WNs. 

\begin{figure}[H]
\centering      
\includegraphics[width=0.5\textwidth]{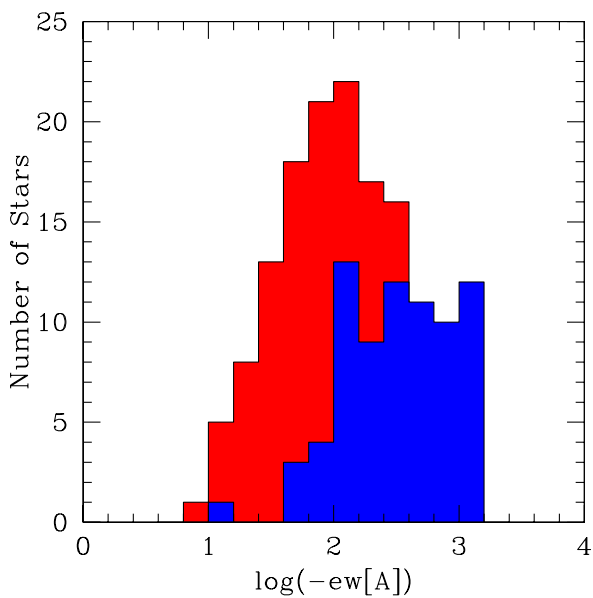}
\caption{\label{fig:WNstrength} Line strengths of galactic and LMC WRs. The red histogram shows the line strengths (measured as the log of the equivalent width) for a WN's strongest emission line, H{\sc ii} $\lambda$4686. The blue histogram shows the same for the WC's strongest emission line, C{\sc iii/iv} $\lambda$4650. The WC's strongest line is up to 4$\times$ stronger than the WN's strongest line making WC stars much easier to detect.}
\end{figure}

The basis for this observational bias is shown in Figure~\ref{fig:WNstrength}. The strongest emission feature in WCs is nearly 4$\times$ stronger than the strongest line in WNs, making WNs much more difficult to detect than WCs of similar brightness \cite{ContiMassey89}.  (More accurately, this is an issue of line fluxes; see treatments in \cite{MJ98} and \cite{Neugent18}.) Thus, while a galaxy (or catalog such as BAT99) might be complete for WC-type stars, there might be a number of missing WNs since their emission lines are so much weaker. The exclusion of these stars would bias the WC to WN ratio to higher values, much like we see when we compare the relative number of WRs observed to that predicted by the Geneva Evolutionary models.  Indeed, this was particularly a problem for M31. The ratio of 2.2 shown in Fig.~\ref{fig:Geneva2005} is the galaxy-wide average for M31, including the older photographic work; if instead one used only the 8 CCD fields, this value would drop to 0.9 \cite{MJ98}, giving strong credence to selection effects being responsible for the problem.

The lack of a galaxy-wide survey for M31 or M33 as well as the possibility of crowding and a strong observational bias against WN stars lead us to conduct our own survey for WRs in M31 and M33.

\section{New Era of Discoveries}
As discussed above, as of 2005, the observed WC/WN ratio was quite poorly aligned with the theoretical predictions at higher metallicities. Thus, M31 and M33 were two ideal regions to study. M31 has the highest metallicity of the Local Group galaxies at log(O/H) + 12 = 8.9 \cite{Zaritsky94,NelsonM31}. M33 has a strong metallicity gradient going from log(O/H) + 12 = 8.3 in the outer regions up to log(O/H)+12 = 8.7 in the inner regions \cite{Magrini2004}.  Thus, these two galaxies presented the perfect opportunity to re-examine the differences between theory and observations.

In 1985 Massey \& Armandroff had pioneered the use of interference filter imaging with CCDs to identify WR candidates \cite{AM85}.  However, the small size of the CCDs available at that time limited the area that could be covered and the large read-noise limited the sensitivity.   An equally large problem, however, was the use of photometry to identify candidates.  This method was far superior to "blinking by eye," as had been used in the photographic studies by \cite{WrayCorso,MS83,MS87}, and allowed "statistically significant" candidates to be identified.  However, the fraction of false positives was overwhelming, simply given the large number of stars involved.  

In the mid-2000s along came large format CCD Mosaic cameras, such as those implemented on the Kitt Peak and Cerro Tololo 4-meter telescopes.  CCDs now had read-noises of 3~e- rather than 100~e-, and these mosaic cameras made it practical to cover all of M31 and M33 in a finite number of fields.  Equally importantly, supernova and transient searches had required the development of the powerful technique of image subtraction, where the the PSFs were matched between two images, and one image subtracted from another to identify images.  We took advantage of both of these improvements in conducting our own searches. 

\subsection{Identification of Candidate WRs}
Searching for candidate WRs was done using the same method in both galaxies as is detailed in \cite{NeugentWRM31, NeugentWRM33}. Overall, the method combines photometric observations using an interference filter system with image subtraction and photometry for candidate detection. 

\begin{figure}[H]
\centering      
\includegraphics[width=0.35\textwidth]{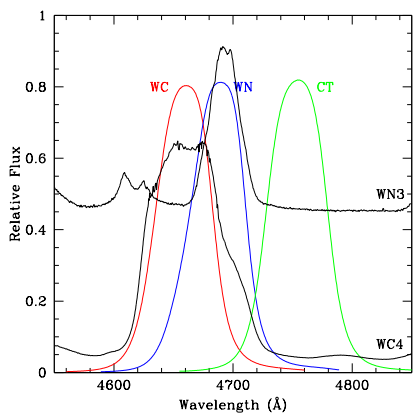}
\caption{\label{fig:bandpass} Filter bandpasses of WN, WC and CT filters. The WN and WC filters are centered on the strongest lines of the WC and WN-type WRs while the CT is centered on the neighboring continuum. Figure adapted from \cite{PaperI}.}
\end{figure}

Thanks to the WR's strong emission lines, they're relatively simple to detect using the appropriately designed interference filters.  Taft Armandroff and Massey used spectrophotometry for WR and non-WR to design a 3-filter system that was optimized identifying WRs in the optical \cite{AM85}.  All three filters have $\sim$ 50\AA\ wide bandpasses, with one centered on the strongest optical line in a WC's spectrum, C{\sc iii/iv} $\lambda$4650 ("WC" filter), another centered on the strongest optical line in a WC's spectrum, He{\sc ii} $\lambda$4686 ("WN" filter) and a third on the neighboring continuum at $\lambda 4750$ ("CT" filter). (Placement of the continuum filter to the red of the emission-line filters is crucial; otherwise, red stars show up as candidates.)   The bandpasses are shown placed atop the spectrum of both an LMC WC- and WN-type WR in Figure~\ref{fig:bandpass}. This filter set was used by \cite{AM85} to search for WRs in the Local Group galaxy dwarfs NGC~6822 and IC~1613, as well as two small test regions of M33. Such work was then extended to selected regions of M33 \cite{Massey98} and M31 \cite{MAC86}, and for the galaxy-wide survey of the SMC \cite{Massey01} discussed above. With these interference filter images in hand, there are two main methods of determining stars that are brighter in the on-band filters (WC and WN) vs.\ in the continuum (CT). The first is using image subtraction and the second is using photometry.

\begin{figure}[H]
\centering      
\includegraphics[width=0.7\textwidth]{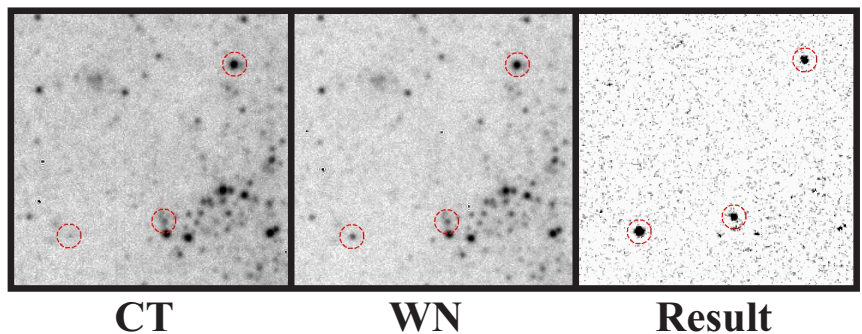}
\caption{\label{fig:subtracted} WR-detection through image subtraction. Three known WRs are outlined in red dashed circles. After subtracting the continuum filter from the WN filter, the resulting image shows three WRs as black stars. This method was used to search for candidate WRs. Figure from \cite{NeugentWRM33}.}
\end{figure}

As mentioned above, image subtraction has been used with great success to detect small brightness changes between on and off band photometry by the supernovae community \cite{Yuan2010}. Simply subtracting the CT from the WC filter should yield candidate WCs while subtracting the CT from the WN filter should yield candidate WNs. However, seeing variability and small changes in pixel scales across the images turn this simple idea into a complex problem and thus cross-convolution methods and point-spread fitting techniques must be used. Example programs include the Astronomical Image Subtraction by Cross-Convolution program \cite{YuanAkerlof08} and High Order Transform of PSF ANd Template Subtraction ({\sc hotpants}) \cite{hotpants}. An example resulting image is shown in Figure~\ref{fig:subtracted} where the background stars have been subtracted out and the candidate WRs are left behind.

As discussed above, most WRs are formed in dense OB associations (in fact, Neugent \& Massey found that 80\% of the WRs in M33 were found in OB associations \cite{NeugentWRM33} with only 2\% being truly isolated). This dictates the need for crowded field photometry to determine the magnitude differences between the WC-CT and WN-CT filters. Armandroff \& Massey had adopted Peter Stetson's {\sc daophot} crowded field photometry software \cite{daophot}, with subsequent modifications and porting to {\sc iraf} \cite{daophotiraf}.  Careful matching in crowded regions must be performed by eye.  Photometry is obtained for all the stars on each on-band exposure (WC, WN), and then matched with the photometry for the same stars on the CT exposure. A zero-point adjustment is then made so that the average difference was zero, and then stars that were more than 3$\sigma$ brighter on either the WC or WN filter exposure when compare to the continuum exposure can be identified.

\subsection{M33}
\begin{figure}[H]
\centering      
\includegraphics[width=0.35\textwidth]{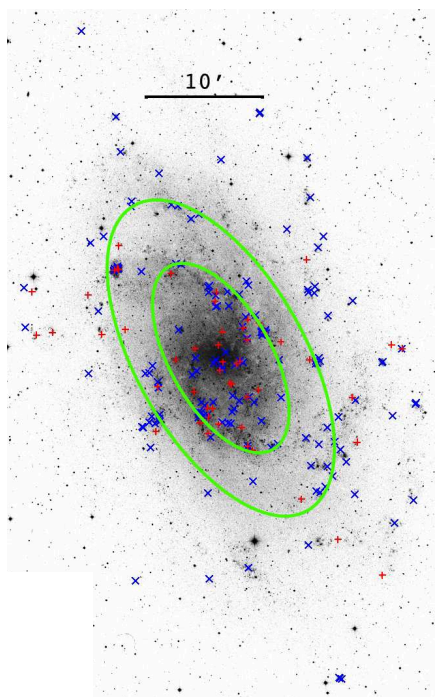}
\caption{\label{fig:M33loc} Location of known WC and WN stars in M33. WN stars are represented as blue $\times$s while WC stars are represented as red +s. The green ovals represent distances of $\rho$ = 0.25 (1.9 kpc) and $\rho$ = 0.50 (3.8 kpc) within the plane of M33. The metallicity gradient extends outward with higher metallicity in the middle and lower in the outer regions. Figure from \cite{NeugentWRM33}.}
\end{figure}

Neugent et al.\ completed the first galaxy-wide survey for WRs using a combination of the image subtraction and photometric method as discussed above \cite{NeugentWRM33}. Overall, they discovered 54 new WRs bringing the total number of confirmed WRs in M33 up to 206, a number they believe is complete to $\sim 5\%$. A majority of these new discoveries were WNs suggesting that the previous WC/WN ratio had been biased towards the easier to find WCs. The locations of the known WRs across the disk of the galaxy are shown in Figure~\ref{fig:M33loc}. Notice that the galaxy has been divided up into three regions representing the strong metallicity gradient with the inner region having a higher metallicity than the outer region.

As discussed in the introduction, the formation of WRs is highly dependent on mass-loss rates which is, in turn, dependent on the metallicity of the environment. In higher metallicity environments, the mass-loss rates will be higher leading to the creation of more WCs. Thus, we expect the WC/WN ratio to be higher in regions of high metallicity, such as in the center of M33. Indeed, this is what we find. While the full comparison of WC/WN ratios vs.\ metallicity will be discussed later, Table~\ref{tab:M33ratio} shows the WC/WN ratio vs.\ metallicity for the inner, middle, and outer regions of M33. (The cut-offs for these regions are a little different than had been used in the earlier study by \cite{MJ98} shown in Figure~\ref{fig:Geneva2005}.)
\begin{table}[H]
\caption{WC/WN ratio vs.\ metallicity for the inner, middle, and outer regions of M33}
\centering
\label{tab:M33ratio}
\begin{tabular}{c c c c c c}
\toprule
\textbf{Region} & \textbf{$\bar{\rho}$} & \textbf{log(O/H) + 12} & \textbf{\# WCs} & \textbf{\# WNs} & \textbf{WC/WN}\\
\midrule
$\rho < 0.25$ & 0.16 & 8.72 & 26 & 45 & $0.58\pm0.09$\\
$0.25 \ge \rho < 0.50$ & 0.38 & 8.41 & 15 & 54 & $0.28\pm0.07$\\
$\rho \ge 0.50 $ & 0.69 & 8.29 & 12 & 54 & $0.22\pm0.06$\\
\bottomrule
\end{tabular}
\end{table}

\begin{figure}[H]
\centering      
\includegraphics[width=0.25\textwidth]{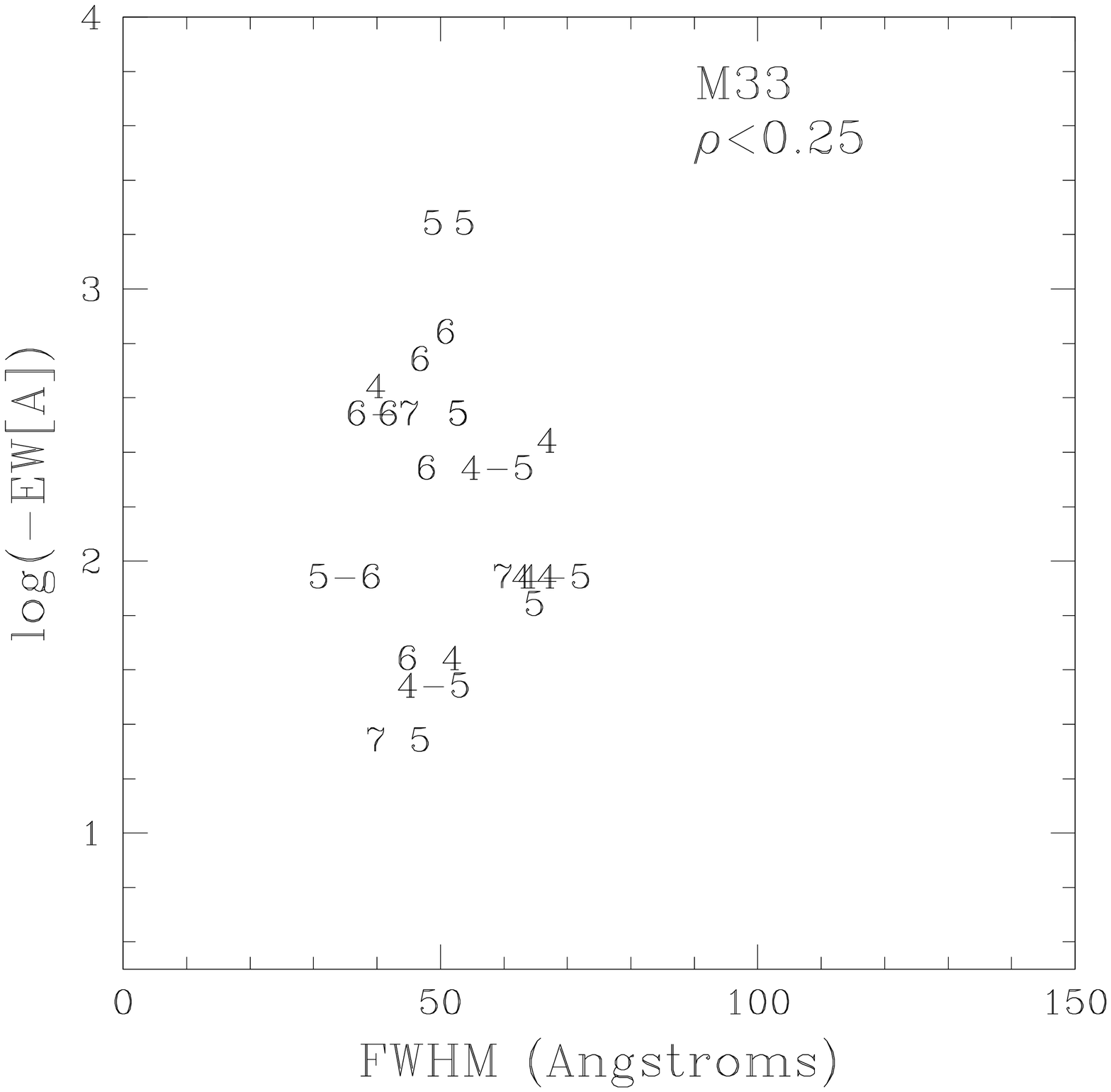}
\includegraphics[width=0.25\textwidth]{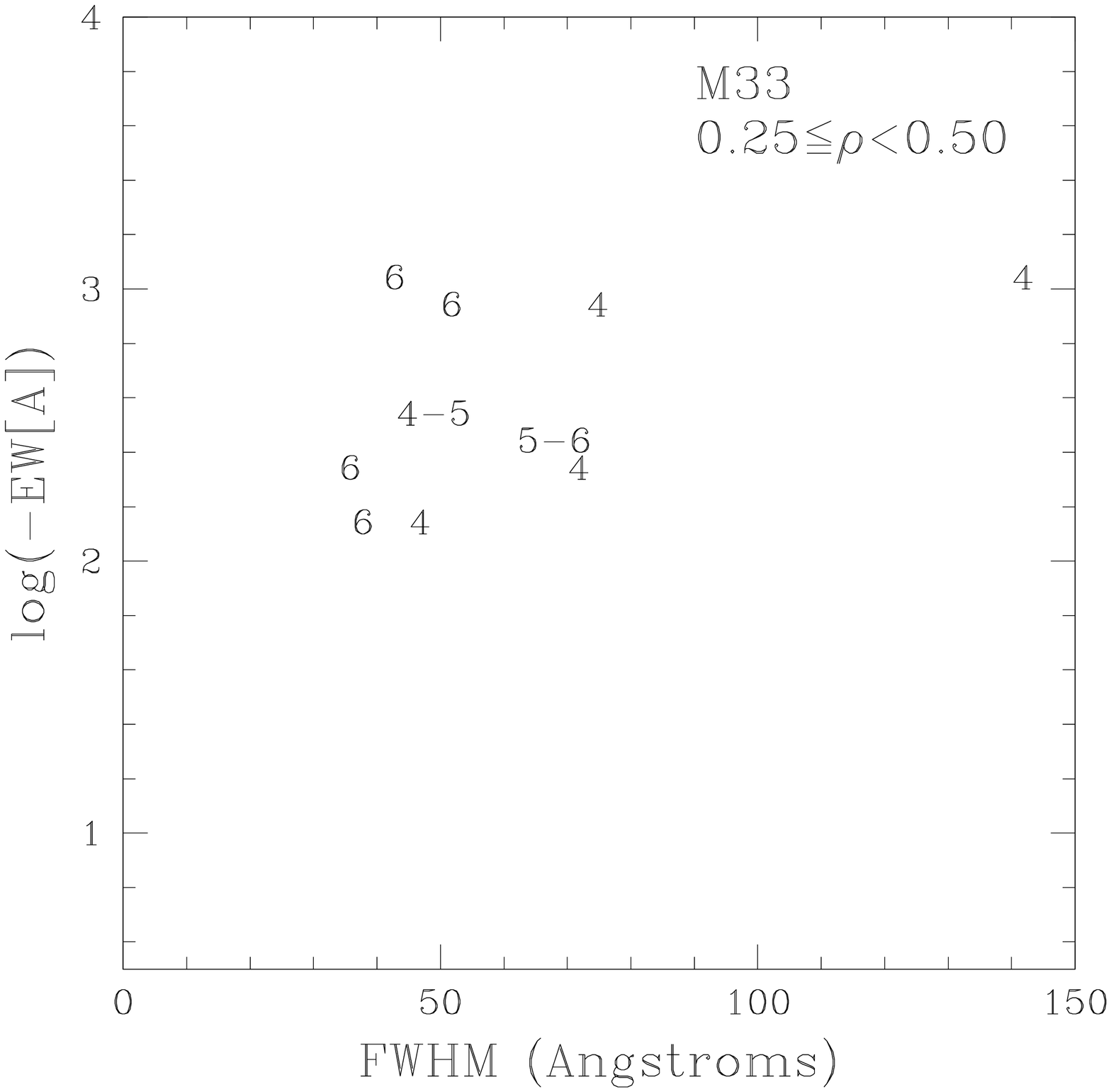}
\includegraphics[width=0.25\textwidth]{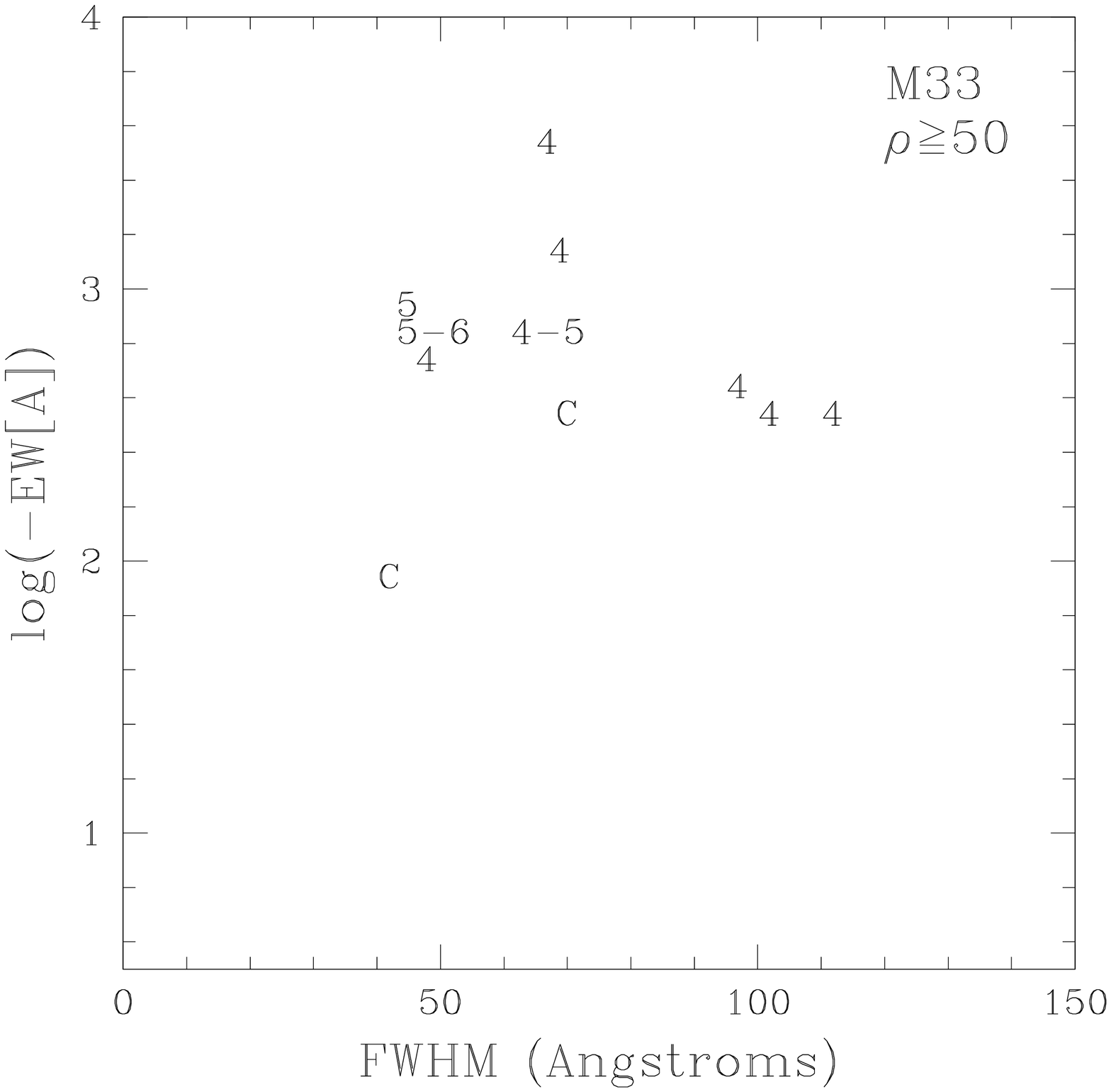}
\caption{\label{fig:hockeystick} WC line strength vs.\ line width. The line strength of the C{\sc iv} $\lambda$5806 line is plotted against its line width with the spectral subtypes indicated (with "C" used if the subtype has not been well established). Notice how the FWHM increases and subtype decreases as the metallicity decreases (larger values of $\rho$). Figure from \cite{NeugentWRM33}.}
\end{figure}

The metallicity gradient of M33 also allows us to probe the relative number of early and late type WCs vs.\ metallicity. Smith first discovered that nearly all of the late-type WCs are found in higher metallicity environments than the early-type WCs \cite{Smith1968}. Additionally, late-type WCs have C{\sc iv} $\lambda5806$ lines that both have smaller equivalent widths and smaller full width half max values than early-type WCs. Thus, plotting these two values against each other vs.\ metallicity shows that the spectral type becomes earlier as metallicity decreases. This is shown in Figure~\ref{fig:hockeystick}. This proves, independent of any direct metallicity measurements, that the metallicity of M33 increases towards the center of the galaxy.

With this new data discussed in Neugent \& Massey, the WC/WN ratio was determined for three regions of medium to high metallicity \cite{NeugentWRM33} and the number of WRs was thought to be complete to 5\%.

\subsection{M31}
\begin{figure}[H]
\centering      
\includegraphics[width=0.5\textwidth]{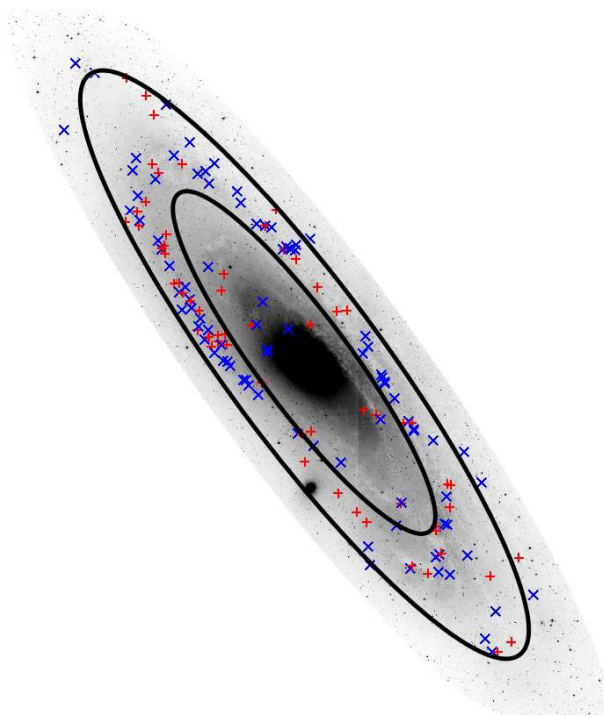}
\caption{\label{fig:M31loc} Locations of all known WR stars in M31. The blue $\times$s represent WN stars while the red +s represent WC stars. The inner black ellipse is at 9 kpc ($\rho$ = 0.43) within the plane of M31, and the outer one is at 15 kpc ($\rho$ = 0.71) which represents the location of the majority of OB forming associations within M31. Figure from \cite{NeugentWRM31}.}
\end{figure}

The next study was done in M31 by Neugent et al. \cite{NeugentWRM31} which has an even higher metallicity than that of the inner region of M33. By using the same detection methods of interference filter imaging, image subtraction and photometry, they discovered 107 new WRs (79 WNs and 28 WCs) bringing the total number of WRs in M31 up to 154, a number they argue is good to within 5\%. They additionally found that 86\% of the observed WRs were within known OB associations as determined by van den Bergh \cite{vandenBergh1964}. The locations of the WRs are shown in Figure~\ref{fig:M31loc}. Due to the addition of the new WNs, the WC/WN ratio dropped from 2.2 down to 0.67. While this helped bring the observations closer to that of the theoretical model predictions, the full story will be told in Section 7. 

Subsequent to the this study, Mike Shara and collaborators discovered an additional WR star in M31, a WN/C star \cite{SharaM31}.  Such objects have WN-like spectra but strong C\,{\sc iv} $\lambda$5806,12 line.  The star is located in strong nebulosity, and is described as heavily reddened (although no specific values are given), and the authors speculate based on this one object that there might be a large population of unfound WRs lying on the "far side" M31's disk, i.e., that only lightly reddened specimens have been found so far.  Is this reasonable?  First we note that the width of the "blue plume" (denote OB stars) in the color magnitude diagram of M31 has a similar width to that of the LMC; compare Figures 10 and 12 in \cite{Massey07}.  If there were a huge population of highly reddened stars we would expect the blue plume to be high asymmetric, with a large tail extending to redder magnitudes.  Secondly, we can do a crude estimate of what we might expect.  We note that the total extinction through the MW's disk is $\sim$0.4~mag in B \cite{deVacext}.  If M31 is similar, then at an inclination of 77$^\circ$ to the line of sight we expect the total extinction in B from one side to the other to be about 1.8~mag, or in V, about 1.4~mag.  This is only 0.6~mag greater than the 0.8~mag in $A_V$ found for OB stars in some of the handful of well-studied OB associations \cite{MAC86}, and is unsurprising.  Thus, although a handful of heavily reddened WRs may certainly have been missed (consistent with the ten that Shara et al.\ estimate), it seems unlikely that there is an opaque wall obscuring WRs on the far side of M31.   

\subsection{Magellanic Clouds}
Thanks to previous surveys, such as the BAT99 catalog, the population of WRs in the MCs was thought to be complete. However, over the years a few unexpected discoveries were made. Perhaps the most surprising of which was of a rare strong-lined WO discovered in the LMC in the rich OB association of Lucke-Hodge 41 \cite{NeugentWO}. Since the BAT99 catalog, six new WRs were discovered before the addition of this new WO suggesting that perhaps our knowledge of the WR content of the LMC was still not complete. Thus, a new search for WRs in the MCs was launched \cite{PaperI,PaperII,PaperIII}. A summary of the results can be found in \cite{Neugent18}.

The overall process of this survey was similar to finding WRs in M31 and M33. The entire optical disks of both the LMC and SMC were observed using the 1-m Swope telescope on Las Campanas, with the three filter interference system and then a combination of image subtraction and photometry was used to detect candidate WRs before they were spectroscopically confirmed.

In the SMC, no new WRs were discovered. However, this isn't too surprising given that there are only 12 known WRs in the entire galaxy \cite{Massey01}, and that the Massey \& Duffy survey had covered the entire galaxy. All of them are of WN type except one binary WO. Further characteristics, such as their physical properties and binary status are discussed later.

The LMC, however, held many surprises. Overall, the new study found 15 new WRs bringing the total number of WRs in the LMC up to 152. Five of them were normal WNs that had been missed due to crowded fields and faint emission lines. However, ten of them were unlike any WR we had seen before.

\begin{figure}[H]
\centering      
\includegraphics[width=0.5\textwidth]{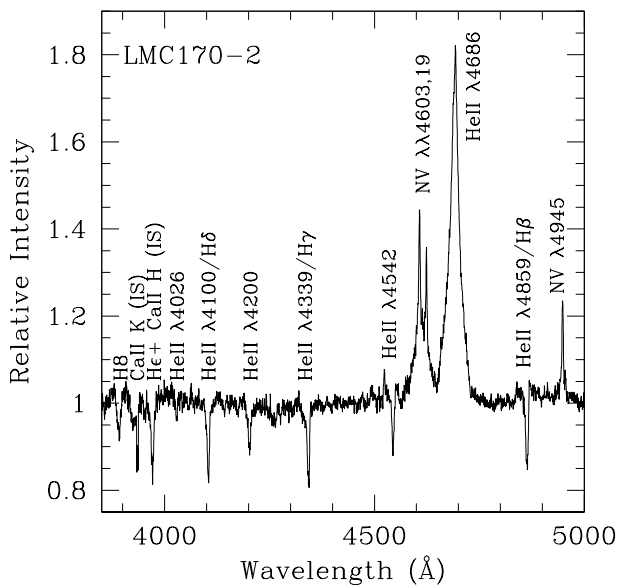}
\caption{\label{fig:WN3O3} Spectrum of LMC170-2, one of our newly discovered WN3/O3 type stars. The WN3 classification comes from the star's N\,{\sc v} emission ($\lambda \lambda$ 4603,19 and $\lambda$4945), but lack of N\,{\sc iv}. The O3 classification comes from the strong He{\sc ii} absorption lines but lack of He{\sc i}. Figure from \cite{NeugentWN3O3}.}
\end{figure}

The spectra of these stars contain absorption lines like that of a O3 star with emission lines like that of a WN3, thus leading to a designation of WN3/O3s \cite{NeugentWN3O3}. A spectrum of one such star showing both the narrow absorption lines and broad emission lines is shown in Figure~\ref{fig:WN3O3}. While their spectra initially suggests binarity, these stars are simply too faint to be WN3 + O3V binaries. The absolute magnitude of an O3V by itself is $M_V \sim -5.5$ while the absolute magnitudes of these WN3/O3s are around $M_V \sim -2.5$. Thus, they could not be in systems with even brighter O3Vs. For this, and other reasons detailed in \cite{NeugentWN3O3}, these stars are single in nature. A further description of their physical parameters and hypothesized place in massive star evolution is discussed in Section 6.

\begin{figure}[H]
\centering      
\includegraphics[width=0.4\textwidth]{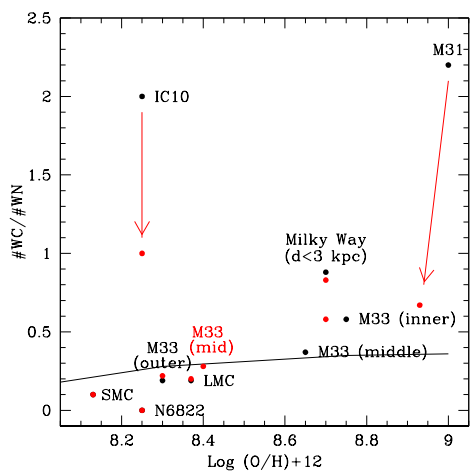}
\caption{\label{fig:wcwncomp} Updated comparison of WC/WN ratio of observed results vs.\ Geneva Evolutionary models. Notice the drastic changes between the old and new values for IC10 and M31. However, the lack of agreement between the models and observations at high metallicities remains.}
\end{figure}

In Figure~\ref{fig:wcwncomp} we now show the effect that the recent work of ourselves and others have made in our knowledge of the WC/WN ratio as a function of metallicity.   Clearly the biggest improvements have come about for M31 and IC10.  However, even for IC 10 the results are still very uncertain, with \cite{MasseyHolmes} finding many additional candidates that have not yet been certified by spectroscopy, and \cite{Tehrani} finding a small number which have also not yet been observed.  For the Milky Way (MW), we took the current 661 in Paul Crowther's on-line catalog, and selected only those with Gaia distances $<$3~kpc using the (model-dependent) catalog of Bailer-Jones et al. \cite{BailerJones}.  This found 99 WRs.  Despite the vast improvement in the distances available since the estimate of the MW's WC/WN by Massey \& Johnson \cite{MJ98}, the value for the WC/WN ratio is essentially unchanged.  Still, as emphasized earlier, construction of a volume-limited sample for the MW is fraught with difficulties.

\section{Wolf-Rayets Beyond the Local Group}

\subsection{Individual WR Populations}

WR stars have been found in a number of more distant galaxies.  NGC~300 is a spiral galaxy in the Sculptor Group (1.9~Mpc) \cite{N300dist}, the nearest galaxy group outside the Local Group.  Broad WR features were found in the spectra of several of NGC300's H\,{\sc ii} regions in the 1980s \cite{Dod300,Deh}.  Eighteen individual WRs were subsequently identified in the early 1990s by interference imaging and follow-up spectroscopy by Gerard Testor, H. Schild, and Breysacher \cite{N300a,N300b,N300d}, with a nineteenth one discovered by accident by Fabio Bresolin and collaborators \cite{N300e}.  A concerted survey with the 8-meter VLT by Schild and collaborators brought this total up to 60, a value which they state is close to complete \cite{N300f}.   Subsequently an additional 9 WRs were found by Crowther and collaborators \cite{N300g}, bringing the total to 69.

How complete do we expect such surveys to be? The distance to NGC 300 is $2.4\times$ larger than the distance to M33, and with similar reddenings, WR stars will be nearly $6\times$ fainter; crowding will be also be 2.4$\times$ larger.  Thus, given what was involved in obtaining a (nearly) complete sample of WRs in M33 by Neugent et al.\ using imaging on a 4-m telescope, one may question how well completeness can be achieved by a telescope only twice as large in aperture.  M33 has 206 WRs.  What would we expect the population to be scaling by the integrated H$\alpha$ luminosities?  The integrated H$\alpha$ luminosity is considered to be one of the "gold standards" of recent star formation activity, and (corrected for reddening and distance) is about 2.1$\times$ greater in M33 than in NGC300 \cite{Rob}. Thus one would naively expect NGC300's WR population to number about 100.  

The most interesting discovery to come out of the NGC~300 studies was Crowther et al.'s discovery that one of the WR stars is coincident with a bright, hard X-ray source \cite{N300g}.  Prior to this, only the Milky Way's Cyg X-3 and IC10-X1 were known as a WR+compact companion (neutron star or black hole) system; see, e.g., discussion and references in \cite{N300g}.  Analysis by Crowther and his team led to a mass of 37$M_\odot$ for the WR star, and $>10M_\odot$ for the compact companion, placing it firmly in the black hole camp.

Other surveys have been carried out for WR stars in even more distant systems with the 8-m VLT by Lucy Hadfield, Joanne Bibby and Crowther: IC 4662 (2.3~Mpc) \cite{4662}, NGC 7793 (3.4~Mpc) \cite{7793}, NGC 1313 (4.1~Mpc) \cite{1313}, M83 (4.5~Mpc) \cite{M83}, NGC 5068 (5-7~Mpc) \cite{5068}, NGC 6744 (7-11~Mpc) \cite{6744}.  Most interesting, perhaps, has been their {\it HST} study
of M101, a large spiral located at a (relatively speaking) modest 6.7~Mpc distance \cite{M101a,M101b}, with followup spectroscopy with the Gemini 8-m \cite{M101c}.  To these we note the more recent study of the WR content of NGC~625 (3.9~Mpc) \cite{625} by integral field spectroscopy on the VLT by Ana Monreal-Ibero and collaborators.

Although these systems are all too far for completeness to be reached to determine the WC/WN ratio reliability, or provide other diagnostics for testing evolutionary models, they are potentially very useful were one of these stars to become a Type Ibc supernova sometime in the near future \cite{WTFc,WTFb,WTFa}.  Thus patience may be required to achieve the scientific benefits of these studies of more distant systems. It is also worth noting that no supernova progenitor has yet to be identified as a WR star \citep{Eldridge2015,Fraser2017}.

\subsection{Integrated WR Populations}
Distant starburst galaxies (by "distant" we means not resolved into stars) often display a WR "bump" in their optical spectra at rest wavelengths of 4650-4670\AA, due to a mixture of WN and WC stars in the integrated spectrum.  The first such system was identified in the compact dwarf He 2-10 \cite{Allen76}; quantitative analysis in theory allows one to derive the relative number of WR and O stars \cite{KunthSargent81}; for a more on this subject, see  \cite{Angel} and other papers in their series.

Kim Sokal and collaborators detected the WR bump in an emerging "super star clusters," massive clusters which are just now clearing out their natal material, demonstrating that the time to clear out such material is comparable to the time it takes for massive stars to evolve to the WR phase ($\sim$3~Myr) \cite{Sokal15,Sokal16}.

\section{Binarity}
One of the most heavily debated questions in massive star research is the issue of binarity. Observations have shown that a significant but still contested fraction of massive stars are found in binary systems. Studies of un-evolved massive stars typically find an observed binary fraction of 30-35\% for O-type stars in relatively short period (less than $\sim$100 days) systems \cite{Garmany80,Sana30Dor}. When long-period systems are included, this percentage approaches 70\% or higher \cite{Gies08,SanaSci}. This question of binarity also extends to WRs. Methods range from light curve analysis, searching for spectral signatures (such as radial velocity variations), and the presence of x-ray emission.  As discussed earlier, the galaxies of the Local Group provide an excellent test-bed for such studies as we are able to determine a complete sample of WRs with which to study the binary fraction.  

Over the decades, many papers have attempted to tackle the issue of binarity head-on. In 1981, Massey \& Conti found that the fraction of Galactic WR stars that were close WR+O star systems was $\sim$25\%, and thus the total fraction must be $<$50\% when the issue of compact companions were included \cite{Massey81}. In 2001, van der Hucht compiled an updated list of WRs in the Galaxy bringing the total up to 227 \cite{vanDer01}. They found that the binary fraction of observed and probable binaries was around 40\%. Foellmi et al. published papers in 2003 looking at the Magellanic Clouds finding close binary fractions of 40\% in the SMC and 30\% in the LMC \cite{Foellmi03SMC, Foellmi03LMC}. More recently, in 2014, Neugent et al.\ obtained multi-epoch spectra of nearly all of the WRs in M31 and M33 and searched for short period binary systems by observing radial velocity variations within the prominent emission and hydrogen absorption lines. Such hydrogen lines tend to suggest the presence of an O-type star companion (with the notable exceptions being the WN3/O3s, and some hydrogen-rich WRs found in the Galaxy and in the SMC) \cite{Neugent14}. This study found that $\sim$30\% of the WRs within M31 and M33 were in short-period binary systems. They additionally found that there was no correlation between binarity and metallicity. Thus, overall, the close binary fraction of WRs appears to be around 30-40\% within all metallicity cases, similar to what is observed for O-type stars. (The exact definition of "close" is a debatable one, but we use here a "spectroscopist's definition," corresponding to detection of orbital motions on the order of several 10s of km s$^{-1}$, corresponding to periods of order 100~days or less for massive stars.)

One further way of searching for WR binaries is through the presence of hard X-ray emission. Most single WRs show soft X-ray emission produced by the winds of the single stars. However, in WR binaries, harder, more luminous X-ray emission forms due to the macroscopic shock interactions between the winds in a binary bound system \cite{Stevens92, Rauw16}. Such X-ray signatures have been found in a few known binary WRs. One of the most extreme such examples is Mk 34 located in the rich OB association of 30 Doradus in the LMC. It has been classified as a WN5ha and is thought to have a (disputably) high mass of 380 M$_\odot$ as derived through spectroscopic analysis \cite{Crowther2011},  but see also \cite{Tehrani2019}.  Garofali et al.\ additionally found a candidate colliding wind binary (WC + O star) in M31 that is located in the dense HII region NGC 604. It is not nearly as bright as Mk 34, but it still shows X-ray emission as discovered by {\it Chandra} \cite{Garofali2019}. While searching for X-ray emission is not the most prominent way of detecting WR binaries, it is more frequently being used as a method of determining binarity.  

As one of our good friend and colleague often reminds us, "One can never prove any star is {\it not} a binary."  That said, another colleagues has noted that the presence of a companion star often makes itself known in the spectrum, albeit in subtle ways.

In single star evolution, the type of WR is heavily influenced by the metallicity of the gas out of which the star formed.  As discussed in the introduction, WN stars that show the hydrogen burning byproducts will appear before WC stars which show the helium burning byproducts. Thus, in a low metallicity environment, one expects to find fewer WCs than in a high metallicity environment. However, once binary evolution is considered, this metallicity dependence decreases because the stripping is being done by Roche-lobe overflow instead of metal-driven stellar winds. Thus, one test of binarity is to look for an excess of WCs in an environment. Or, even more compelling, is to identify the even more evolved WOs (oxygen-rich WRs) in low metallicity environments. There are two prime examples of such stars that were most likely created through binary evolution. The first is the WO star in the SMC. As discussed earlier, there are only 12 known WRs in the SMC (a low metallicity environment of $0.25\times$ solar) and 11 of them are WNs, as expected. However, the 12th one is a WO that should only form in a high metallicity environment \cite{Moffat1983,Moffat1990}. There is an additional example of a WO forming in the low metallicity environment of IC1613 \cite{Tramper2019}, which has a metallicity of $\sim0.15\times$ solar \cite{Talent1980}. Although evolution to the WO stage is not expected by even the most massive single stars in low metallicity environments, models that include binary evolution do predict WOs in low metallicity environments \cite{BPASS2.1}. These two stars are thus examples of WRs likely forming through binary evolution;  doubtlessly there are many more.

While many studies have shown the close binary fraction to be around 30-40\%, the actual value is still hotly debated. Proponents of binary evolution argue that the currently single WR stars were once multiple, but their companions have merged.  There is little evidence, however, to support this conjecture. There is additionally the question of whether the WRs that formed from binary evolution began with initial masses great enough to suggest that they would have become WRs anyway and the binary mechanism simply sped up the process. Thus, it is possible that the importance of binary evolution may be somewhat overstated, even if the fraction of WRs in binary systems is higher than currently observed.

\section{Physical Parameters}
As is characteristic of stars approaching the Eddington Limit, a WR's spectrum is heavily influenced by strong stellar winds and high mass-loss rates \cite{Grafener2011}. Keeping the model's luminosity near, but below, the Eddington limit can make modeling WRs quite a challenge. Additionally, the stars' high surface temperatures mean that the assumption of local thermodynamic equilibrium (LTE) is no longer valid. Instead, the high degree of ionization (and correspondingly decreased opacity) causes the radiation field to decouple from the local thermal field. Furthermore, WR atmospheres are significantly extended when compared to their radius. Thus, plane-parallel geometry cannot be used, and instead spherical geometry must be included.  The emission lines that characterize WR spectra are produced in the outflowing winds, with mass-loss rates of order $10^{-5} M_\odot$ yr$^{-1}$.  Finally, WR models must be fully blanketed and include the effects of thousands of overlapping metal lines, which occur at the (unobservable) short wavelengths ($<1000$\AA) where most of the flux of the star is produced. Two codes are currently capable of including these complexities: the Potsdam Wolf-Rayet Models, or PoWR \cite{Grafener02}, and the CoMoving Frame GENeral spectrum analysis code, {\sc cmfgen} \cite{CMFGEN}. For a much more detailed description of the physics and complexities involved in modeling a WR, see e.g., \cite{Crowther07,Hillier15,Sander17}.

There have been few modeling campaigns of complete samples of WRs in galaxies other than the Magellanic Clouds. In M31, for example, 17 late-type WNs were modeled using PoWR in an attempt to learn more about the wind laws of such stars in different metallicity environments \cite{Sander2014}.  One limitation of this study was the lack of UV spectroscopy.  Nevertheless, they were able to place luminosity constraints on the modeled WRs for values between $10^5$ and $10^6$ L$_\odot$ and suggest that WRs in M31 form from initial mass ranges between 20 and 60 $M_\odot$. This is similar to that found in both the Galaxy and Magellanic Clouds. However, no modeling has taken place for the WC stars in a high metallicity environment like M31.

Conversely, much modeling has been done of WRs in the Magellanic Clouds. Over the past few years, surveys of single and binary WNs in both the SMC and LMC, and the WN3/O3 stars in the LMC have all been performed. In 2014, Hainich et al.\ determined physical parameters of over 100 WNs in the the LMC using grids of PoWR \cite{Hainich14} models. They concluded that the bulk ($\sim$ 88\%) of the WRs analyzed had progressed through the RSG before becoming WRs thus implying that they evolved from 20-40 M$_\odot$ progenitors. They also found that these results were well aligned with studies of Galactic WRs suggesting that there is no metallicity dependence on the range of main sequence masses that evolve into WRs. This research in the LMC was extended to the WR binaries by Shenar et al.\ in 2019 \cite{Shenar2019}, who looked at the 44 binary candidates and found that 28 of them have composite spectra and 5 of them show periodically moving WR primaries. They conclude that while $45\pm30$\% of the WNs in the LMC have most likely interacted with a companion via mass-transfer, many of these WRs would have evolved to become WRs through single star evolution. 

Both the binary and single WNs in the SMC have also been modeled using the PoWR code \cite{Hainich15, Shenar2016}. As discussed earlier, many of the WNs in the SMC have absorption lines that, if not due to a companion, could simply be photospheric lines that are inherent to the stars because of their weak stellar winds. Thus, studying them for photometric and radial velocity variability is necessary to determine their binarity. Based on modeling with the PoWR code, it was concluded again that while some of these stars are binaries now, they still would have become WRs through single-star evolution given their high initial main-sequence masses.

As discussed above, there has been additional modeling of the LMC WN3/O3s using {\sc cmfgen}. All ten of these stars show strong absorption and emission lines as is shown in Figure~\ref{fig:WN3O3} for one of the newly discovered stars. {\sc cmfgen} spectral line fitting was used to determine the physical parameters of these ten stars. Table~\ref{tab:WN3O3s} shows the range of values for the 10 WN3/O3s compared to typical values for an O3V and WN3 star in the LMC (WN3 parameters from \cite{Hainich14}. O3V parameters from \cite{Massey13}.). While the temperature is a bit on the high side for what we would expect for a LMC WN, the majority of the parameters are within the expected ranges. The one exception is the mass-loss rate which is more similar to that of an O3V than of a normal LMC WN.

\begin{table}[H]
\caption{Physical parameters of WN3/O3s, WNs, and O3Vs in the LMC}
\centering
\label{tab:WN3O3s}
\begin{tabular}{c c c c}
\toprule
& \textbf{WN3/O3s} & \textbf{WN3} & \textbf{O3V}\\
\midrule
$T_{\rm off}$ (K) & 100,000 - 105,000 & 80,000 & 48,000 \\
$\log \frac{L}{L_\odot}$ & 5.6 & 5.7 & 5.6 \\
$\log \dot{M}$ & -6.1 - -5.7 & -4.5 & -5.9 \\
He/H (by \#) & 0.8 - 1.5 & 1.0-1.4 & 0.1 \\
N (by mass) & 5-10$\times$ solar & 5-10$\times$ solar & 0.5$\times$ solar \\
M$_V$ & -2.5 & -4.5 & -5.5\\
\bottomrule
\end{tabular}
\end{table}

Although other WN stars with intrinsic absorption lines are known, the WN3/O3s appear to be unique \cite{NeugentWN3O3, Shenar2019}, and their place in the evolution of massive stars still unknown. Neugent's study \cite{NeugentWN3O3} considered the possibility that these stars were the products of homogenous evolution, a situation that can occur if the star is rotating so rapidly that mixing keeps the composition nearly uniform within the star (see, e.g., \cite{LamersLevesque}).  However, they ruled this out based upon the stars' low rotational velocities combined with low mass-loss rates, as the latter implies that the high angular momentum could not have been carried off by stellar winds. Based on their absolute magnitudes they are not WN + O3V binaries, though they could be hiding a less-massive companion. It is additionally possible that binarity influenced their previous evolution. However, it is currently thought that instead these stars represent an intermediate stage between O stars an WNs. More research is ongoing in an attempt to answer this question.

\section{Comparisons to Evolutionary Models}
As discussed in the Introduction, comparing the observed WC/WN ratio with evolutionary model predictions is one of the most important reasons to search for WRs. Currently we have complete samples of the WR populations for the Magellanic Clouds, M31, and M33. The galaxy's metallicities and WC/WN ratios are shown in Table~\ref{tab:WRratio}.  We have included the Milky Way, although here the data are far less certain that the statistical uncertainties would indicate. As is expected, the WC/WN ratio increases with increasing metallicity due to the strength of the stellar winds. We can now compare these observational results to those of the evolutionary models.

\begin{table}[H]
\caption{WC/WN ratio vs.\ metallicity for the Local Group Galaxies}
\centering
\label{tab:WRratio}
\begin{tabular}{c c c c c c}
\toprule
\textbf{Region} & \textbf{log(O/H) + 12} & \textbf{\# WCs and WOs} & \textbf{\# WNs} & \textbf{WC/WN}\\
\midrule
SMC & 8.13 & 1 & 11 & $0.09\pm0.09$\\
M33 outer & 8.29 & 12 & 54 & $0.22\pm0.06$\\
LMC & 8.37 & 28 & 124 & $0.23\pm0.01$\\
M33 middle & 8.41 & 15 & 54 & $0.28\pm0.07$\\
Milky Way &  8.70 & 46 &53 & $0.83\pm0.10$\\
M33 inner & 8.72 & 26 & 45 & $0.58\pm0.09$\\
M31 & 8.93 & 62 & 92 & $0.67\pm0.11$\\
\bottomrule
\end{tabular}
\end{table}

There are two primary sets of evolutionary models currently used in the massive star community. The first is the Geneva Evolutionary Models \cite{Meynet05} that model the evolution of single stars. The other is the Binary Population and Spectral Synthesis (BPASS) models that focus on binary evolution \cite{Eldridge2008,Eldridge2016}. Besides the obvious difference between the two of modeling single vs.\ binary stars, the models also have some important differences. In the case of the Geneva models, there are only results for a few metallicities, as is shown in the figure below. This makes comparisons between the observations and the models quite difficult because there are only a few points. However, these models have been created with different initial rotation rates as it plays quite a large effect on the resulting physics. Conversely, the BPASS models have results spanning a wide range of metallicities, but these models do not include rotation. So, due to these differences, it is difficult to compare the observations directly to either set of models. However, in time, the models will continue to improve.

\begin{figure}[H]
\centering      
\includegraphics[width=0.5\textwidth]{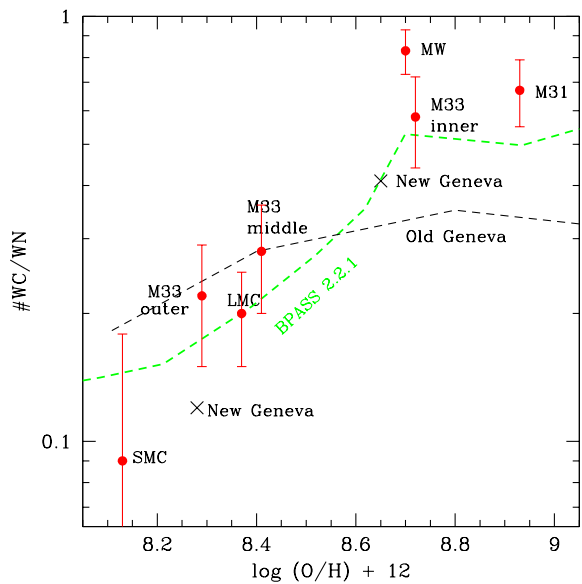}
\caption{\label{fig:bpass} WC/WN ratio vs.\ metallicity compared to both BPASS2.2.1 and Geneva Evolutionary models. Notice the improved results between the observed WC/WN ratio and the Geneva Evolutionary models, but the lack of models at a variety of metallicities. Also notice the good agreement between the BPASS2.2.1 models and both the observed results. The error bars come from $\sqrt{N}$ statistics; see \cite{MJ98,NeugentWRM31}.}
\end{figure}

In Figure~\ref{fig:bpass} we show the agreement between the WC/WN ratios and the evolutionary models.   We have not included NGC~6822 or IC~1613 in this diagram, as they each have too few WRs for meaningful statistics (4 and 1, respectively).  We also have not included IC~10, as we feel the current value is, at best, an upper limit.  We have included the value for the MW determined as described above, although we suspect that this too is an upper limit.  As for the predictions: The solid line is from the older Geneva evolutionary models, the first to include rotation \cite{Meynet05}.  The green dashed line is an updated version of the predictions from BPASS2 \cite{BPASS2.1}, and these 2.2.1 predictions were kindly provided by J. J. Eldridge (2019, private communication).  The models assume continuous star formation, a Salpeter IMF slope, and an upper mass limit of 300$M_\odot$.  The BPASS models also include the effects of binary evolution.  Finally, the two $\times$'s denote results from the latest single-star evolutionary models.  The higher metallicity value comes from \cite{Cyril12}, while the lower metallicity point was computed by Cyril Georgy from preliminary Geneva z=0.006 models, and used in \cite{NeugentWRM31}. There is good agreement between the newer Geneva single-star models and the binary evolution models; this may simply be that the BPASS models do not yet include the effects of rotation.  Including rotation can reduce the expected ratio of WC/WN stars; see Figure 10 in \cite{NeugentWRM31}.  Although the observational data at all metallicities are now in relatively good shape, improvements are still pending in the evolutionary models.  Still, we can conclude that the large issue at high metallicity with the oldest models has largely gone away.

\section{Summary and the Future of WRs}
WRs are the bare stellar cores of massive stars, and the last stage in a massive star's lifetime before they turn into supernovae. Observing a complete set of both the nitrogen and carbon rich WRs within a galaxy allows for important comparisons between the observed WC/WN ratio and that predicted by the evolutionary models. Because the evolution of WRs is highly dependent on the metallicity of the surrounding environment, it is important to do these comparisons across a wide range of galaxies with different metallicities, such as the galaxies in the Local Group.

Finding WRs observationally is done using a combination of interference filters and photometric techniques before the identified candidates are confirmed spectroscopically. This method has been used with great success over the past few decades and lead to the discovery of hundreds of WRs in both our galaxy and even those far enough away that we can only observe the integrated light coming from clusters of WRs. While this method has lead to the discovery of mostly complete samples of WRs within the Local Group, there is still much progress to be made in more distant galaxies.

The binary fraction of WRs is still highly contested with current observations putting it somewhere between 30-40\% for the close binary frequency. However, as the distance between binaries expands, and the effect of binarity on the evolution of WRs in the past is considered, it is difficult to fully understand what role binaries play in the evolution of WRs. Modeling the spectra of the currently known WRs using sophisticated modeling codes such as PoWR and {\sc cmfgen} allow us to get a better handle on the physical properties of both the binaries and single stars and compare them across a wide range of metallicities. 

As discussed in Section 2, while much progress has been made in the field of WR research, there is still much to be done. With {\it Gaia} it is now possible to determine distances to nearby WRs within our own Galaxy leading to better observations of their reddenings and better modeling of their physical properties. We are additionally learning more about the content of other types of massive stars (such as O/B stars, RSGs, etc.) that allow us to compare the ratio of those stars vs.\ WRs to the evolutionary model predictions placing further constraints on the models. Finally, we are continuing to push the observational boundaries to further and further galaxies in an attempt to observe complete samples of WRs in both the galaxies of the Local Group and beyond!

\funding{This research was partially funded by the National Science Foundation, most recently through AST-1612874, as well as through Lowell Observatory.} 

\acknowledgments{The authors acknowledge all of their dear friends, family, and collaborators who have supported them though many years of Wolf-Rayet research. They additionally thank J.J. Eldridge for her help better understanding WR binary evolution, Paul Crowther for useful information on Galactic WR surveys, as well as an anonymous referee whose suggestions improved the paper.}

\conflictsofinterest{The authors declare no conflict of interest.} 

\externalbibliography{yes}
\bibliography{masterbib.bib}

\end{document}